Corresponding Author: Khaled Qandalji,

Corresponding Author's Institution:

First Author: Khaled Qandalji

Order of Authors: Khaled Qandalji

Manuscript Region of Origin:

Abstract:
   In this article, we carry out the Hamiltonization in the axial gauge ,of the t'Hooft-Polyakov monopole field outside the localized region, which represents the monopole's core. One feature of the treatment here, is using the Higgs vacuum condition as both strong and weak equation instead of using it in the degree of freedom reduction.

# Dirac Quantization of t'Hooft-Polyakov Monopole Field: Axial Hamiltonization

K. Rasem Qandalji

*Amer Institute*
*P.O. Box 1386, Sweileh, 11910*
*JORDAN*
*E-mail:* qandalji@hotmail.com

## ABSTRACT

In this article, we carry out the Hamiltonization in the axial gauge, of the t'Hooft-Polyakov monopole field outside the localized region, which represents the monopole's core. One feature of the treatment here, is using the Higgs vacuum condition as both strong and weak equation instead of using it in the degree of freedom reduction.

## 1. Introduction

The t'Hooft-Polyakov monopole model [1] consists of an *SO*(3) gauge field interacting with an isovector Higgs field $\phi$, whose non-singular extended solution looks, at large distances, like a Dirac monopole.

The model's Lagrangian is:

$$\mathcal{L} = -\frac{1}{4} G_a^{\mu\nu} G_{a\mu\nu} + \frac{1}{2} D^\mu \phi . D_\mu \phi - V(\phi)$$

where:

$$\phi = (\phi_1, \phi_2, \phi_3), \quad \text{and} \quad V(\phi) = \frac{1}{4}\lambda(\phi_1^2 + \phi_2^2 + \phi_3^2 - a^2)^2$$

$G_a^{\mu\nu}$ is the gauge field strength: $\quad G_a^{\mu\nu} = \partial^\mu W_a^\nu - \partial^\nu W_a^\mu - e\varepsilon_{abc} W_b^\mu W_c^\nu$, where $W_a^\mu$ is the gauge potential.

i

Let the monopole configuration be centered at the origin, the requirement of total energy finiteness implies that there is some radius $r_0$ such that for $r \geq r_0$ we have, to a good approximation:

$$D^\mu \boldsymbol{\phi} \equiv \partial^\mu \boldsymbol{\phi} - e\mathbf{W}^\mu \times \boldsymbol{\phi} = 0 \tag{1}$$

$$\phi_1^2 + \phi_2^2 + \phi_3^2 - a^2 = 0, \quad (\Rightarrow V(\boldsymbol{\phi}) = 0). \tag{2}$$

Regions of space-time, where the above equations are satisfied, constitute the Higgs Vacuum.

The symmetry group $SO(3)$, generated by $T_a$'s, is spontaneously broken, by the Higgs Vacuum, down to $U(1)$ generated by $\dfrac{\boldsymbol{\phi}.\mathbf{T}}{a}$.

The general form of $\mathbf{W}^\mu$ satisfying (1), provided $\boldsymbol{\phi}$ satisfies (2), is[2]:

$$\mathbf{W}^\mu = \frac{1}{a^2 e}\boldsymbol{\phi} \times \partial^\mu \boldsymbol{\phi} + \frac{1}{a}\boldsymbol{\phi} A^\mu, \tag{3}$$

where $A^\mu$ is arbitrary.

It follows that:

$$\mathbf{G}^{\mu\nu} = \frac{1}{a}\boldsymbol{\phi} F^{\mu\nu} \tag{4}$$

where,

$$F^{\mu\nu} = \frac{1}{a^3 e}\boldsymbol{\phi}.(\partial^\mu \boldsymbol{\phi} \times \partial^\nu \boldsymbol{\phi}) + \partial^\mu A^\nu - \partial^\nu A^\mu \tag{5}$$

So in Higgs vacuum, $\mathcal{L}$ will reduce to:

$$\mathcal{L} = -\frac{1}{4} G_a^{\mu\nu} G_{a\mu\nu},$$

and on account of (2) and (4), we get:

$$\mathcal{L} = -\frac{1}{4} F^{\mu\nu} F_{\mu\nu} \tag{6}$$

[We will use the metric(+,-,-,-).]

## 2. Hamiltonization

To quantize a theory canonically, we need first to hamiltonize it, that is to find the Hamiltonian describing the system as a function of the dynamical variables and their conjugate momenta only. Finding such a Hamiltonian is easy only in the standard case, in which the conjugate momenta are independent functions of the velocities. This is not the case here: Our conjugate momenta are not all independent and we will have to apply the Dirac algorithm for constrained systems [3],[4].

In the monopole field region, where (1) and (2) are satisfied, i.e. in the Higgs Vacuum, $\mathcal{L}$ is given by:

$$\mathcal{L} = -\frac{1}{4} F^{\mu\nu} F_{\mu\nu} = -\frac{1}{4}\left[ \frac{1}{a^6 e} \varepsilon_{ijk}\varepsilon_{rst}\phi_i\phi_r \partial^\mu \phi_j \partial^\nu \phi_k \partial_\mu \phi_s \partial_\nu \phi_t \right.$$

$$\left. + 2(\partial^\mu A^\nu - \partial^\nu A^\mu)\partial_\mu A_\nu + \frac{4}{a^3 e}\varepsilon_{ijk}\phi_i \partial^\mu \phi_j \partial^\nu \phi_k \partial_\mu A_\nu \right]. \qquad (7)$$

The Conjugate momentum of dynamical coordinates $\phi_\ell(\mathbf{x})$ is:

$$\pi_\ell(x) \equiv \frac{\partial \mathcal{L}}{\partial \dot{\phi}_\ell(x)} =$$

$$= \frac{\varepsilon_{ij\ell}}{a^3 e} \phi_i \partial^k \phi_j \left( \frac{\varepsilon_{rst}}{a^3 e}\phi_r \partial_0 \phi_s \partial_k \phi_t + \partial_0 A_k - \partial_k A_0 \right). \qquad (8)$$

The conjugate momentum of dynamical coordinates, $A^\eta(\mathbf{x})$ is:

$$\Pi_\eta(x) \equiv \frac{\partial \mathcal{L}}{\partial \dot{A}^\eta(x)} = \frac{\varepsilon_{rst}}{a^3 e}\phi_r \partial_\eta \phi_s \partial_0 \phi_t + \partial_\eta A_0 - \partial_0 A_\eta$$

$$= \begin{cases} 0 & ,\text{for } \eta = 0 \\ F_{i0} & ,\text{for } \eta = i = 1,2,3 \end{cases} \qquad (9)$$

By comparing (8) with (9), we arrive at the following relations between the momentum variables: $\quad \pi_\ell(\mathbf{x}) = -\frac{\varepsilon_{ij\ell}}{a^3 e}\phi_i(\mathbf{x})\partial^k \phi_j(\mathbf{x})\Pi_k(\mathbf{x}), \quad$ where $\ell = 1,2,3$.

So we get the "primary" constraints:

$$\Phi_\ell(\mathbf{x}) \equiv \pi_\ell(\mathbf{x}) + \frac{\varepsilon_{ij\ell}}{a^3 e}\phi_i(\mathbf{x})\partial^k \phi_j(\mathbf{x})\Pi_k(\mathbf{x}) \approx 0, \qquad \text{where } \ell = 1,2,3 \qquad (10)$$

and

$$\Phi_0(\mathbf{x}) \equiv \Pi_0(\mathbf{x}) \approx 0, \qquad (11)$$

and since we are restricting our region to the Higgs Vacuum, we also impose the strong condition (2) as a constraint:

$$\chi(\mathbf{x}) \equiv \phi_1^2(\mathbf{x}) + \phi_2^2(\mathbf{x}) + \phi_3^2(\mathbf{x}) - a^2 \approx 0 \quad (2a)$$

( (2a) will be used as a strong equation whenever possible, despite it being incorporated into the formulation as a weak equation as well.) Using (11), we can solve for $\dot{A}^i$ on the constraint surface, call it $\bar{\dot{A}}^i$:

$$\bar{\dot{A}}^i(\phi_\ell, A^n, \Pi_j, \dot{\phi}_k) = \frac{\varepsilon_{rst}}{a^3 e} \phi_r \partial_0 \phi_s \partial_i \phi_t + \Pi_i - \partial_i A^0 \quad (12)$$

On the constraint surface, the Hamiltonian density, $H$, is equal to function of the coordinates and momenta, call it $\mathcal{H}$ ,[3],[4],[5] where:

$$\mathcal{H} \equiv \left[ \left( \partial \mathcal{L} / \partial \dot{\phi}_\ell \right) \dot{\phi}_\ell + \left( \partial \mathcal{L} / \partial \dot{A}^n \right) \dot{A}^n - \mathcal{L} \right]_{\dot{A}^i = \bar{\dot{A}}^i} =$$

$$= \frac{1}{2} \Pi_i \Pi_i - \Pi_i \partial_i A_0 + \frac{1}{2} \partial^i A^j \left( \partial_i A_j - \partial_j A_i \right) + \frac{\varepsilon_{rst}}{a^3 e} \phi_r \partial^i \phi_s \partial^j \phi_t \partial_i A_j$$

$$+ \frac{\varepsilon_{ijk} \varepsilon_{rst}}{4 a^6 e^2} \phi_i \phi_r \partial^m \phi_j \partial^n \phi_k \partial_m \phi_s \partial_n \phi_t$$

$$= \frac{1}{2} \Pi_i \Pi_i - \Pi_i \partial_i A_0 + \frac{1}{4} F_{ij} F_{ij} \quad (13)$$

Now, using Eqs. (13), (10), (11), and (2a), we find that the consistency conditions[6]: $\dot{\Phi}(q,p) \approx 0$, will lead to one new "secondary" constraint, namely:

$$\dot{\Phi}_0 = \Pi_{0,0} = \int d^3 x' \{\Pi_0(\mathbf{x}), H(\mathbf{x}')\} = \int d^3 x' \{\Pi_0(\mathbf{x}), \mathcal{H}(\mathbf{x}')\} = \partial^i \Pi_i \approx 0 \quad (14)$$

[where the fact that $\Phi_0$ has vanishing Poisson Brackets with other constraints has been used.] $\partial^i \Pi_i$, has identically vanishing Poisson Brackets with $\mathcal{H}$ and all the constraints, and therefore will not lead to any new constraints.

On the other hand, we find there are two independent combinations of $\Phi_1$, $\Phi_2$, $\Phi_3$ and $\chi$ which are first class constraints:

Any combination of the form, $\quad \eta_k \equiv \varepsilon_{ijk} \phi_j \Phi_i - \frac{1}{2} \alpha_k \chi, \quad$ where $k = 1, 2, 3$

and (where, $\alpha_k \equiv \frac{3}{a^3 e} \Pi_\ell \partial^\ell \phi_k$ ), will have vanishing Poisson Brackets with $\Phi_1, \Phi_2$, $\Phi_3$ and $\chi$, on the constraint surface, and therefore with any combinations of them.

On account of $\chi$ being strong equation, (i.e. $\phi_i \partial^\mu \phi_i = 0$), we see that: $\phi_k \eta_k = 0$, and therefore only two of the three $\eta_k$'s are independent. Since $\eta_k$'s and combinations of them are the only possible forms of first class constraints formed from $\Phi_1, \Phi_2, \Phi_3$, and $\chi$.( Allowing combinations that involves, also, (11) and (14) will not help in finding any new independent first class constraints, since (11) and (14) are already first class.) Therefore we can only have two first class constraints formed of $\Phi_i$'s and $\chi$: $\eta_3$ and $\eta_1$ say.

We will replace the set of constraints $\Phi_1, \Phi_2, \Phi_3$, and $\chi$ by $\zeta_1, \zeta_2, \zeta_3$, and $\zeta_4$:

$$\zeta_1 \equiv \eta_3 = \phi_2 \Phi_1 - \phi_1 \Phi_2 - \frac{\alpha_3}{2} \chi$$
$$\zeta_2 \equiv \eta_1 = \phi_3 \Phi_2 - \phi_2 \Phi_3 - \frac{\alpha_1}{2} \chi$$
$$\zeta_3 \equiv \frac{1}{2a^2}\left(\phi_1 \Phi_1 + \phi_2 \Phi_2 + \phi_3 \Phi_3\right)$$
$$\zeta_4 \equiv \chi = \phi_1^2 + \phi_2^2 + \phi_3^2 - a^2$$
(15)

Consistency conditions associated with $\zeta_1$ and $\zeta_2$ will be weakly satisfied on account of $\chi$ being strong equation, (i.e. $\phi_i \partial^\mu \phi_i = 0$), and that $\zeta_1, \zeta_2$ are first class:

$$\int d^3x' \{\eta_k(\mathbf{x}), H(\mathbf{x'})\} \approx$$
$$\approx \varepsilon_{ijk} \phi_j(x) \int d^3x' \{\Phi_i(\mathbf{x}), \mathcal{H}(\mathbf{x'})\} - \frac{1}{2}\alpha_k(x)\int d^3x' \{\chi(\mathbf{x}), \mathcal{H}(\mathbf{x'})\}$$
$$= -\frac{3}{2}\varepsilon_{ijk}\varepsilon_{irs}\phi_j \partial^m \phi_r \partial^n \phi_s F_{mn} + 0$$
$$= 3\phi_k \partial^m \phi_k \partial^n \phi_k F_{mn} = 0, \quad \text{(we used, } \phi_i \partial^\mu \phi_i = 0 \text{,in the step before last.)}$$

Consistency conditions associated with $\zeta_3$ and $\zeta_4$ will not lead to new constraints either, but they will impose conditions on the velocities. It may be worth mentioning at this point, that choosing $\zeta_3$ as above is convenient, since under a certain canonical transformation in which $\zeta_4$ is a coordinate, $\zeta_3$ will be its corresponding conjugate momentum ( unique up to additional terms with weakly vanishing Poisson Bracket with $\zeta_4$.)

The constraint (11), $\Phi_0 \equiv \Pi_0$, is primary first class, and therefore a degeneracy of the Hamiltonian will be associated with it [4], (i.e., solutions of the Lagrangian equations contain an arbitrary function of time associated with $\Phi_0$.)

We lift the above degeneracy by imposing a gauge given by the supplementary condition, (a constraint), call it $\zeta(\mathbf{x})$:

$$\zeta(\mathbf{x}) \equiv A^0(\mathbf{x}) \approx 0. \tag{16}$$

Imposing (16) will lead to contradiction upon passing to quantum theory [6]. Following Dirac, the degree of freedom, $A^0$, will be discarded, because $A^0$ and $\Pi_0$ are restricted to be zero at all time, and therefore they are of no interest to us.

$\mathcal{H}$, (13), upon the above reduction of degrees of freedom, will reduce to:

$$\mathcal{H} = \frac{1}{2}\Pi_i \Pi_i + \frac{1}{4} F_{ij} F_{ij}. \tag{13a}$$

Constraint (14), is first class, and we will call it, $\zeta_5$:

$$\zeta_5 \equiv \partial^i \Pi_i \tag{17}$$

Now, we have three first class constraints: $\zeta_1$, $\zeta_2$ and $\zeta_5$. We, also, have two second class constraints: $\zeta_3$ and $\zeta_4$. Similar to what was done in the case of the constraint, $\Phi_0$, we will impose three supplementary conditions, "gauges", to lift the degeneracy caused by $\zeta_1$, $\zeta_2$ and $\zeta_5$ being first class. The gauge fixing conditions we will impose are:

$$\begin{aligned}\zeta_6 &\equiv \frac{1}{ae}\left(\phi_2 \partial^3 \phi_1 - \phi_1 \partial^3 \phi_2\right) - A^3 \phi_3 \approx 0 \\ \zeta_7 &\equiv \frac{1}{ae}\left(\phi_3 \partial^3 \phi_2 - \phi_2 \partial^3 \phi_3\right) - A^3 \phi_1 \approx 0 \\ \zeta_8 &\equiv A^3 \approx 0\end{aligned} \tag{18}$$

It is clear that $\zeta_8$ is the axial gauge associated with $A^i$'s. Similarly, $\zeta_6$ and $\zeta_7$ are the axial gauge associated with $\mathbf{W}^\mu$, i.e. $\mathbf{W}^3 \approx 0$. From Eq. (3), we can easily see that:

$$\zeta_6 = -\frac{1}{a}\left(\mathbf{W}^3\right)_3 \equiv -\frac{1}{a}W_3^3$$

$$\zeta_7 = -\frac{1}{a}\left(\mathbf{W}^3\right)_1 \equiv -\frac{1}{a}W_1^3.$$

[Notice that we don't need to impose additional constraint to ensure that $\left(\mathbf{W}^3\right)_2 \approx 0$, since this is identically satisfied on the constraint surface, where $\zeta_6$ and $\zeta_7$ are valid, since we have:
$$\mathbf{W}^3.\partial^3\boldsymbol{\phi} = W_a^3\partial^3\phi_a = 0,$$
which we arrive at using Eq. (3), and that $\chi$ is a strong equation (i.e., $\phi_i\partial^\mu\phi_i = 0$).]

The Poisson Brackets amongst the constraints, including the gauge fixing conditions, are given on the constraint surface by the matrix, $C(\vec{x},\vec{x}')$, where:

$$C_{ij}(\mathbf{x},\mathbf{x}') \equiv \{\zeta_i(\mathbf{x}),\zeta_j(\mathbf{x}')\}\Big|_{\zeta_k \approx 0,\ k=1,\ldots,8} \tag{19}$$

After calculating the Poisson Brackets, and then evaluating them on the constraint surface, the non-vanishing elements of the matrix, $C$, will be:

$$C_{16}(\mathbf{x},\mathbf{x}') = -C_{61}(\mathbf{x}',\mathbf{x}) = -\frac{1}{ae}[\phi_1(\mathbf{x})\phi_1(\mathbf{x}') + \phi_2(\mathbf{x})\phi_2(\mathbf{x}')]\partial^{3'}\delta^3(\mathbf{x}-\mathbf{x}')$$

$$C_{17}(\mathbf{x},\mathbf{x}') = -C_{71}(\mathbf{x}',\mathbf{x}) = \frac{1}{ae}\phi_1(\mathbf{x})\phi_3(\mathbf{x}')\partial^{3'}\delta^3(\mathbf{x}-\mathbf{x}')$$

$$C_{18}(\mathbf{x},\mathbf{x}') = -C_{81}(\mathbf{x}',\mathbf{x}) = -\frac{1}{ae}\delta^3(\mathbf{x}-\mathbf{x}')\partial^3\phi_3(\mathbf{x})$$

$$C_{26}(\mathbf{x},\mathbf{x}') = -C_{62}(\mathbf{x}',\mathbf{x}) = \frac{1}{ae}\phi_1(\mathbf{x}')\phi_3(\mathbf{x})\partial^{3'}\delta^3(\mathbf{x}-\mathbf{x}')$$

$$C_{27}(\mathbf{x},\mathbf{x}') = -C_{72}(\mathbf{x}',\mathbf{x}) = -\frac{1}{ae}[\phi_2(\mathbf{x})\phi_2(\mathbf{x}') + \phi_3(\mathbf{x})\phi_3(\mathbf{x}')]\partial^{3'}\delta^3(\mathbf{x}-\mathbf{x}')$$

$$C_{28}(\mathbf{x},\mathbf{x}') = -C_{82}(\mathbf{x}',\mathbf{x}) = -\frac{1}{ae}\delta^3(\mathbf{x}-\mathbf{x}')\partial^3\phi_1(\mathbf{x})$$

$$C_{34}(\mathbf{x},\mathbf{x}') = -C_{43}(\mathbf{x}',\mathbf{x}) = -\delta^3(\mathbf{x}-\mathbf{x}')$$

$$C_{36}(\mathbf{x},\mathbf{x}') = -C_{63}(\mathbf{x}',\mathbf{x}) = \frac{1}{ae}[\phi_2(\mathbf{x})\phi_1(\mathbf{x}') - \phi_1(\mathbf{x})\phi_2(\mathbf{x}')]\partial^{3'}\delta^3(\mathbf{x}-\mathbf{x}')$$

$$C_{37}(\mathbf{x},\mathbf{x}') = -C_{73}(\mathbf{x}',\mathbf{x}) = \frac{1}{ae}[\phi_2(\mathbf{x})\phi_3(\mathbf{x}') - \phi_3(\mathbf{x})\phi_2(\mathbf{x}')]\partial^{3'}\delta^3(\mathbf{x}-\mathbf{x}')$$

$$C_{56}(\mathbf{x},\mathbf{x}') = -C_{65}(\mathbf{x}',\mathbf{x}) = \phi_3(\mathbf{x}')\partial^3\delta^3(\mathbf{x}-\mathbf{x}')$$

$$C_{57}(\mathbf{x},\mathbf{x}') = -C_{75}(\mathbf{x}',\mathbf{x}) = \phi_1(\mathbf{x}')\partial^3\delta^3(\mathbf{x}-\mathbf{x}')$$

$$C_{58}(\mathbf{x},\mathbf{x}') = -C_{85}(\mathbf{x}',\mathbf{x}) = -\partial^3\delta^3(\mathbf{x}-\mathbf{x}')$$

On the constraint surface, (in particular using $\zeta_6$ and $\zeta_7$ combined with $\zeta_8$ and that $\phi_i \partial^\mu \phi_i = 0$), the non-vanishing elements of the inverse matrix, $C^{-1}$, will be:

$$C_{43}^{-1}(\mathbf{x},\mathbf{x}') = -C_{34}^{-1}(\mathbf{x}',\mathbf{x}) = -\delta^3(\mathbf{x}-\mathbf{x}')$$

$$C_{61}^{-1}(\mathbf{x},\mathbf{x}') = -C_{16}^{-1}(\mathbf{x}',\mathbf{x}) = \frac{ae}{\phi_2^2(\mathbf{x})}\left[\frac{1}{a^2}\phi_1^2(\mathbf{x})-1\right]F(\mathbf{x},\mathbf{x}')$$

$$C_{62}^{-1}(\mathbf{x},\mathbf{x}') = -C_{26}^{-1}(\mathbf{x}',\mathbf{x}) = \frac{-ae}{\phi_2^2(\mathbf{x})}\left[\frac{1}{a^2}\phi_1(\mathbf{x})\phi_3(\mathbf{x})\right]F(\mathbf{x},\mathbf{x}')$$

$$C_{65}^{-1}(\mathbf{x},\mathbf{x}') = -C_{56}^{-1}(\mathbf{x}',\mathbf{x}) = \frac{1}{\phi_2^2(\mathbf{x})}\left[\frac{1}{a^2}\phi_1(\mathbf{x})\{\phi_1(\mathbf{x})\phi_3(\mathbf{x}')-\phi_3(\mathbf{x})\phi_1(\mathbf{x}')\}\right.$$
$$\left.-\phi_3(\mathbf{x}')+\phi_3(\mathbf{x})\right]F(\mathbf{x},\mathbf{x}')$$

$$C_{71}^{-1}(\mathbf{x},\mathbf{x}') = -C_{17}^{-1}(\mathbf{x}',\mathbf{x}) = \frac{-ae}{\phi_2^2(\mathbf{x})}\left[\frac{1}{a^2}\phi_1(\mathbf{x})\phi_3(\mathbf{x})\right]F(\mathbf{x},\mathbf{x}')$$

$$C_{72}^{-1}(\mathbf{x},\mathbf{x}') = -C_{27}^{-1}(\mathbf{x}',\mathbf{x}) = \frac{ae}{\phi_2^2(\mathbf{x})}\left[\frac{1}{a^2}\phi_3^2(\mathbf{x})-1\right]F(\mathbf{x},\mathbf{x}')$$

$$C_{75}^{-1}(\mathbf{x},\mathbf{x}') = -C_{57}^{-1}(\mathbf{x}',\mathbf{x}) = \frac{1}{\phi_2^2(\mathbf{x})}\left[\frac{1}{a^2}\phi_3(\mathbf{x})\{\phi_1(\mathbf{x}')\phi_3(\mathbf{x})-\phi_3(\mathbf{x}')\phi_1(\mathbf{x})\}\right.$$
$$\left.-\phi_1(\mathbf{x}')+\phi_1(\mathbf{x})\right]F(\mathbf{x},\mathbf{x}')$$

$$C_{81}^{-1}(\mathbf{x},\mathbf{x}') = -C_{18}^{-1}(\mathbf{x}',\mathbf{x}) = \frac{-ae}{\phi_2^2(\mathbf{x})}\phi_3(\mathbf{x})F(\mathbf{x},\mathbf{x}')$$

$$C_{82}^{-1}(\mathbf{x},\mathbf{x}') = -C_{28}^{-1}(\mathbf{x}',\mathbf{x}) = \frac{-ae}{\phi_2^2(\mathbf{x})}\phi_1(\mathbf{x})F(\mathbf{x},\mathbf{x}')$$

$$C_{85}^{-1}(\mathbf{x},\mathbf{x}') = -C_{58}^{-1}(\mathbf{x}',\mathbf{x}) = \frac{-1}{\phi_2^2(\bar{x})}\left[\phi_1(\mathbf{x})\phi_1(\mathbf{x}')+\phi_3(\mathbf{x})\phi_3(\mathbf{x}')-a^2\right]F(\mathbf{x},\mathbf{x}')$$

[where, we have:

$$\partial^3 F(\mathbf{x},\mathbf{x}') = -\delta^3(\mathbf{x}-\mathbf{x}'),$$

and hence, we get:

$$F(\mathbf{x},\mathbf{x}') = \frac{1}{2}\delta(x^1-x'^1)\delta(x^2-x'^2)\varepsilon(x^3-x'^3),$$

where,

$$\varepsilon(x^3-x'^3) \equiv \text{algebraic sign of } (x^3-x'^3).]$$

## 3. Conclusion

Now, that we arrived at $C^{-1}(\mathbf{x}, \mathbf{x}')$, we can use it to evaluate the Dirac Bracket for arbitrary functions of the coordinates and the momenta, where the Dirac Bracket between $\eta(q(\mathbf{x}), p(\mathbf{x}))$ and $\xi(q(\mathbf{x}'), p(\mathbf{x}'))$, is given by [5],[6]:

$$\{\eta(\mathbf{x}), \xi(\mathbf{x}')\}_{D(\zeta)} \equiv \{\eta(\mathbf{x}), \xi(\mathbf{x}')\} -$$
$$\iint \{\eta(\mathbf{x}), \zeta_\alpha(\mathbf{x}'')\} d^3x'' C^{-1}_{\alpha\alpha'}(\mathbf{x}'', \mathbf{x}''') d^3x''' \{\zeta_{\alpha'}(\mathbf{x}'''), \xi(\mathbf{x}')\}$$

where, $\alpha, \alpha' = 1, 2, ..., 8$, and where $\zeta_\alpha$'s are given by Eqs. (15),(17), and (18).

For computing the Dirac Bracket, second class constraints (i.e., all the constraint available in the theory at this point; the original ones along with the gauge fixing ones), can be treated as strong equations.

To quantize the above "Hamiltonized" classical theory, we have to follow the standard procedure [5], [6]:

(I) Classical variables will correspond to operators acting on the Hilbert space.

(II) Dirac Bracket will correspond to the commutator multiplied by $\frac{-i}{\hbar}$.

(III) The constraint equations are strong relations among operators.

## Acknowledgment

I thank the I&B-Foundation for their generous and continuous support, and I thank Prof. Sudarshan for his help with this work.